\documentclass[submitting]{nst}

\usepackage{subfigure,dcolumn}
\usepackage[T2A,T1]{fontenc}
\usepackage[english]{babel}

\usepackage{ wasysym }
\usepackage{ upgreek }
\usepackage{ amssymb, amsmath }

\usepackage{color}

\begin{document}

\title{
Study of $^{7}$Li+{$^{10}$B} elastic scattering and the lithium-induced reaction of one-nucleon transfers from $^{10}$B($^{7}$Li,$^{6}$Li)$^{11}$B}\thanks{The research was funded by the Russian Science Foundation (project № 24-22-00117) and by the NRC “Kurchatov Institute”.The authors thank Dr. Yuri Sereda and the team of experimental setup COMBAS for their assistance in preparing the experiment.}

\author{Sergey Stukalov}
\affiliation{Joint Institute for Nuclear Research, Dubna, Moscow Region 141980, Russia}

\author{Yuri Sobolev}
\email[Corresponding author 1, ]{Yuri Sobolev, sobolev@jinr.ru}
\affiliation{Joint Institute for Nuclear Research, Dubna, Moscow Region 141980, Russia}
\affiliation{National Research Centre Kurchatov Institute, Moscow 123182, Russia}

\author{Yuri Penionzhkevich}
\affiliation{Joint Institute for Nuclear Research, Dubna, Moscow Region 141980, Russia}
\affiliation{National Research Nuclear University MEPhI, Moscow 115409, Russia}

\author{Nassurlla Burtebayev}
\affiliation{Institute of Nuclear Physics of the Ministry of Energy of the Republic of Kazakhstan, Almaty 050032, Republic of Kazakhstan}
\affiliation{Al-Farabi National University, Almaty 050032, Republic of Kazakhstan}

\author{Sergey Goncharov}
\email[Corresponding author 2, ]{Sergey Goncharov, gsa@srd.sinp.msu.ru}
\affiliation{M.V. Lomonosov Moscow State University, Moscow 119991, Russia}

\author{Yuri Gurov}
\affiliation{Joint Institute for Nuclear Research, Dubna, Moscow Region 141980, Russia}
\affiliation{National Research Nuclear University MEPhI, Moscow 115409, Russia}

\author{Andrey Danilov}
\affiliation{National Research Centre Kurchatov Institute, Moscow 123182, Russia}

\author{Alla Demyanova}
\affiliation{National Research Centre Kurchatov Institute, Moscow 123182, Russia}

\author{Sergey Dmitriev}
\affiliation{National Research Centre Kurchatov Institute, Moscow 123182, Russia}

\author{Maulen Nassurlla}
\affiliation{Institute of Nuclear Physics of the Ministry of Energy of the Republic of Kazakhstan, Almaty 050032, Republic of Kazakhstan}
\affiliation{Al-Farabi National University, Almaty 050032, Republic of Kazakhstan}

\author{Viktar Starastsin}
\affiliation{National Research Centre Kurchatov Institute, Moscow 123182, Russia}
\affiliation{National Research Nuclear University MEPhI, Moscow 115409, Russia}

\author{Alexey Shakhov}
\affiliation{Joint Institute for Nuclear Research, Dubna, Moscow Region 141980, Russia}

\author{Semyon Raidun}
\affiliation{National Research Centre Kurchatov Institute, Moscow 123182, Russia}

\author{Nguyen Hoai Chau}
\affiliation{Joint Institute for Nuclear Research, Dubna, Moscow Region 141980, Russia}
\affiliation{Institute for Science and Technology of Energy and Environment, Hanoi 122000,
Vietnam}

\begin{abstract}
 The angular distributions of elastic scattering of $^{7}$Li, as well as the lithium-induced reaction of one-nucleon transfers $^{10}$B($^{7}$Li,$^{6}$Li)$^{11}$B were measured at $E_\text{lab}$ = 58 MeV. Experiment was done using U-400 accelerator beam of the FLNR JINR, Dubna. Angular distribution for reaction $^{10}$B($^{7}$Li,$^{6}$Li)$^{11}$B with excitation of the 3.56 MeV state ($^{6}$Li*) is presented for the first time. The DWBA analysis of the differential cross section of the $^{10}$B($^{7}$Li,$^{6}$Li)$^{11}$B ground state (g.s.) transition and excited ($J^\pi$ = 0$^+$, $T$ = 1, $E$ = 3.56 MeV) state of $^{6}$Li transition was performed. The optical model potentials were obtained by fitting of measured elastic scattering data and evaluating parameters for the output reaction channels. Phenomenological approach based on solving an approximate equation for the reaction form factor was used to determine its radial dependence and empirical values of asymptotic normalization coefficient (ANC). Obtained values of ANC’s for the $^{6}$Li$_\text{g.s.}$ and $^{6}$Li*(3.56 MeV) states are in agreement with the literature ones. Comparison of the radial dependences of form factors shows that the wave function of the $^{6}$Li nucleus in excited ($J^\pi$ = 0$^+$, $T$ = 1, $E$ = 3.56 MeV) state has increased spatial dimension compared to the ground state. This result is an argument in favor of a halo existence in $^{6}$Li*(3.56 MeV) state, while the question of a halo in $^{6}$Li$_\text{g.s.}$ still leaves open.
\end{abstract}


\maketitle

\section{Introduction}

Experimental studies of the properties of neutron-rich radioactive nuclei located far from the stability line led to the discovery of a new nuclear structure in some nuclei ($^{6}$He, $^{11}$Li) - a neutron halo \cite{TANIHATA1985}. According to modern concepts, such nuclei consist of a dense nuclear core and a low-density outer region formed by valence weakly bound neutrons. Initially, the term "exotic nuclei" itself referred to these nuclei. One of the main manifestations of properties of halo is an increased radius of the valence nucleons' density distributions, 2-3 times greater than the size of the core.

Among nuclei with a two-neutron halo \cite{Riisager2013}, the so-called Borromean structure is quite common - a system, in which each pair of the three participants does not form a bound state, while all three participants form a bound nuclear system that is stable with respect to neutron emission (for example, $^{4}$He+$n$+$n$ and $^{9}$Li+$n$+$n$ in the nuclei of $^{6}$He and $^{11}$Li, respectively).

One of the most interesting issues in the study of exotic nuclei is the problem of preserving the properties of the halo in excited states of nuclei that are isobaric analogues of the ground states of nuclei with a halo. From this point of view, one of the interesting objects, most accessible for experimental study, is the stable nucleus $^{6}$Li, which, together with the nuclei $^{6}$He and $^{6}$Be, forms an isobaric triplet $A$ = 6. 

The ground state of the $^{6}$Li nucleus, according to the hypothesis proposed in \cite{Izosimov2016,Izosimov2017, Izosimov2018, Izosimov2020}, has a “tango structure”, i.e. a quasi-molecular structure “$\alpha$-core” - “deuteron” in which $p$ and $n$ valence nucleons move in a correlated manner in the deuteron, without forming $\alpha$+$p$ and $\alpha$+$n$ bound states with the “$\alpha$-core”.

An argument in favor of this hypothesis can be the energy dependence of the total cross sections $\sigma_\text{R}(E)$ of the reactions $^{6}$He,$^{6,7}$Li+$^{28}$Si \cite{Sobolev2005, Lukyanov2008}, measured at $E_\text{LAB}$ = (7\textcolor{red}{--}20) AMeV, which showed that the values of $\sigma_\text{R}$($^{6}$He,$^{6}$Li+$^{28}$Si) exceed the cross sections of the reactions $^{7}$Li+$^{28}$Si. 

An additional argument for the presence of a halo structure in the ground state of $^{6}$Li can also be the widths of the momentum distributions of $^{4}$He during the breakup of $^{6}$He and $^{6}$Li nuclei \cite{Kalpakchieva2007}. The width of the momentum distribution of $^{4}$He formed in the breakup reaction of $^{6}$Li was $\sim$ 50 MeV/c \cite{Kalpakchieva2007} in standard deviation and occupied an intermediate value between the widths of the $^{4}$He distributions $\sim$ 28 MeV/c and $\sim$ 100 MeV/c obtained in the breakup reactions of $^{6}$He and stable nuclei, respectively.

To date, experimental data on the spatial structure of the excited state of $^{6}$Li ($J^\pi$ = 0$^+$, $T$ = 1, $E$ = 3.56 MeV), which is an isobaric analog of $^{6}$He g.s., are limited. Most of the studies \cite{CortinaGil1996, CortinaGil1998, Brown1996, Li2002, Galanina2014}, investigated multi-step reaction channels, ($t$, $p$), charge exchange $^{1}$H($^{6}$He,$^{6}$Li)$n$, etc. Analysis of multi-step reaction channels requires consideration of a larger number of possible combinations \cite{Galanina2014}, which complicates obtaining unambiguous results, in contrast to stripping and pickup reactions. 

Based on the results of the study of the charge exchange reaction $^{1}$H($^{6}$He,$^{6}$Li)$n$ \cite{CortinaGil1996, CortinaGil1998, Brown1996, Li2002}, a conclusion was made about the two-nucleon halo structure of the state $^{6}$Li ($J^\pi$ = 0$^+$, $T$ = 1, $E$ = 3.56 MeV), similar to $^{6}$He$_\text{g.s.}$. Also in the work \cite{Demyanova2018} a comparison of the $rms$-radii of the ground ($^{6}$He) and excited ($^{6}$Li*) isobar-analog states was carried out, based on the analysis of differential cross sections obtained in different experiments. An estimate of the $rms$-radius of the state $^{6}$Li($J^\pi$ = 0$^+$, $T$ = 1, $E$ = 3.56 MeV) \cite{Demyanova2018} was made within the framework of the Modified Diffraction Model method \cite{Danilov2009} in the analysis of the angular distributions of the cross sections of elastic and inelastic scattering of $^{3}$He+$^{6}$Li \cite{Givens1972}. The obtained radius, within the error limits, coincided with the calculations within the ab initio NCSM (no-core shell model) \cite{Rodkin2023}. Moreover, the calculations \cite{Rodkin2023} showed that the radius of the ground state of $^{6}$Li, within the error limits, coincides with the radius of the 3.56 MeV state of $^{6}$Li.

At present, there are insufficient experimental data to draw an unambiguous conclusion about the spatial structure of $^{6}$Li$_\text{g.s.}$. Experimental values of the matter radius of $^{6}$Li$_\text{g.s.}$ vary significantly within the range from $r_\text{m}$ = 2.09 $\pm$ 0.02 fm \cite{Tanihata1985a} to $r_\text{m}$ = 2.45 $\pm$ 0.07 fm \cite{Dobrovolsky2006}. 

It should be noted that until now there were no experimental results on direct transfer reactions, of which $rms$-radii of $^{6}$Li$_\text{g.s.}$  and excited state of $^{6}$Li($J^\pi$ = 0$^+$, $T$ = 1, $E$ = 3.56 MeV) (the isobaric-analog state of $^{6}$He$_\text{g.s.}$) can be obtained simultaneously. 

The aim of this work is the simultaneous measurement of angular distributions of differential cross sections of the transfer reaction $^{10}$B($^{7}$Li,$^{6}$Li)$^{11}$B with transition to the ground $^{6}$Li$_\text{g.s.}$ and excited $^{6}$Li($J^\pi$ = 0$^+$, $T$ = 1, $E$ = 3.56 MeV) states together with channels of elastic scattering $^{7}$Li+$^{10}$B and, based on the analysis of these data, to obtain information about the spatial structure of the $^{6}$Li nucleus states, and the quantitative characteristics of the interaction in the decay vertex $^{7}$Li $\rightarrow$ $^{6}$Li+$n$ (asymptotic normalization coefficients) in these states of the $^{6}$Li nucleus.

\section{Experimental setup}

\begin{figure*}[htb]
\includegraphics
  [width=0.9\hsize]
  {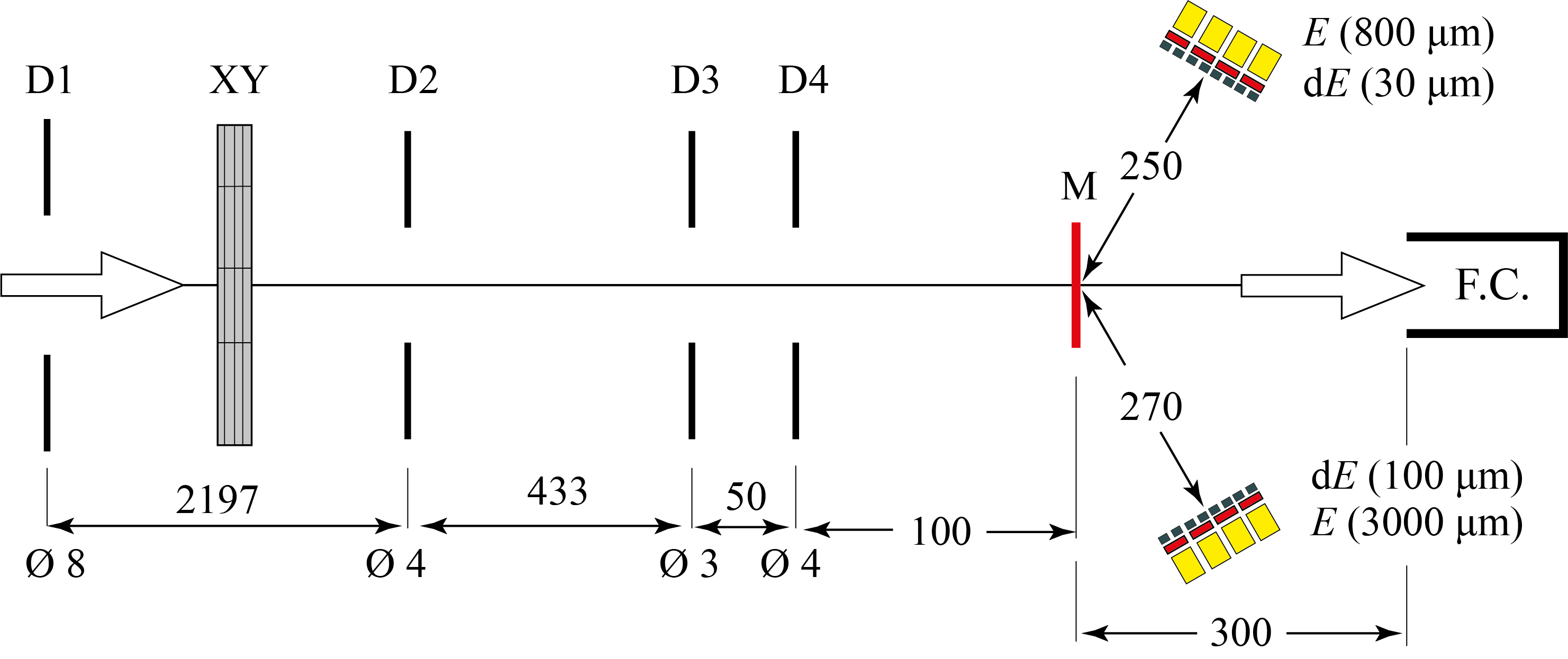}
\caption{Schematic view diagram of the experimental setup for measuring differential cross sections of $^{7}$Li+$^{10}$B reaction products. The values of diameters and distances between the setup elements are given in millimeters.}
\label{fig:01}
\end{figure*}

Measurement of angular distributions of differential cross sections of $^{7}$Li+$^{10}$B reaction products was performed using the U-400 cyclotron of the JINR Laboratory of Nuclear Reactions, Dubna. Fig. \ref{fig:01} shows a schematic diagram of the experimental setup elements. The $^{7}$Li ion beam with energy $E$ = 58 MeV and energy resolution $\Delta \text{E}_\text{(FWHM)}$= 0.5 MeV was focused using a system of magnets in the ion guide channel, a position-sensitive multiwire “X-Y” chamber, and was formed by diaphragms D1-D4, which formed a beam collimator (see Fig. \ref{fig:01}). The beam formed by the collimator in the target position M had a beam spot diameter $\diameter$ = 3.6 mm and an angular aperture of $\Updelta \Uptheta \approx 0.2^\circ$.

A self-supporting target of the isotope $^{10}$B ($t$ $\approx$ 0.05 mg/cm$^2$, $\diameter$ = 1.0 cm), containing impurities of the isotopes $^{12}$C (9\%) and $^{16}$O (6\%), was installed in the center of the reaction chamber perpendicular to the beam axis.

The impurities were estimated using known experimental data on elastic scattering in the reactions $^{7}$Li+$^{10}$B \cite{Etchegoyen1988}, $^{7}$Li+$^{12}$C \cite{Schumacher1973}, $^{7}$Li+$^{16}$O \cite{Schumacher1973} at close energies ($E_\text{lab}$ = 36 MeV and 39 MeV).

At a distance of 30 cm from the target M in beam direction a Faraday cup F.C. was located. (Fig. \ref{fig:01}), which is a thick-walled ($\sim$0.5 cm) steel pipe with a diameter of 4 cm and a height of 25 cm. The beam current data measured by Faraday cup and the beam integrator unit (ORTEC – 439) was recorded together with the experimental data by the data acquisition system (DAQ). 

Additional control of the beam and the state of the target $M$ was carried out using a monitor $E$ Si(Li)-detector (not shown in Fig. \ref{fig:01}).

The DAQ of the experimental setup consisted of the VME electronics units (MVLC crate controller \cite{mesytec}, MADC-32 ADC \cite{mesytec}) and NIM standard logical units and operated by MVME Mesytec program \cite{mesytec}. 

The DAQ master trigger was the MVLC VME block, to the input of which signals of a request to record an event were received from any of the detectors of the setup, as well as pulses from the beam current integrator unit. The DAQ dead time values were estimated by the number of MVLC input request pulses and the number of events written to files. The DAQ system was used to measure the angular distributions of the differential cross sections of the reaction products, which were recorded using two detector groups (see Fig. \ref{fig:01}). Each detector group consisted of four $\Delta E - E$ telescopes of Si detectors. 
The first group of $\Delta E$-$E$ telescopes contains two pairs of $\Delta E$-$E$ telescopes and was used to measure particles emitted in the forward angles ($7^\circ < \Uptheta < 30^\circ$). Each pair of telescopes of the first group consists of two $\Delta E$ detectors (100 $\upmu$m thick and 12x12 mm$^{2}$ of sensitive area) and one $E$ Si(Li) detector (3000 $\upmu$m thick and diameter $D$=20 mm). The second group of $\Delta E$-$E$ telescopes was used to detect particles emitted at angles $\Uptheta > 17^\circ$. The four telescopes of the second group consisted of one $\Delta E$ detector (30 $\upmu$m thick and $D$=20 mm in diameter) and an 800 $\upmu$m thick strip detector located behind it. The sensitive area of each strip was equal to 6x50 mm$^{2}$.

In front of each telescope, there were lead diaphragms of square cross section $\sim$3×4 mm$^2$, providing a solid angle of the detectors of the telescopes of the first and second groups $\Omega = 1.3\times 10^{-4}$ sr and $2.2\times 10^{-4}$ sr, respectively. 

\begin{figure*}[htb]
\includegraphics
  [width=0.9\hsize]
  {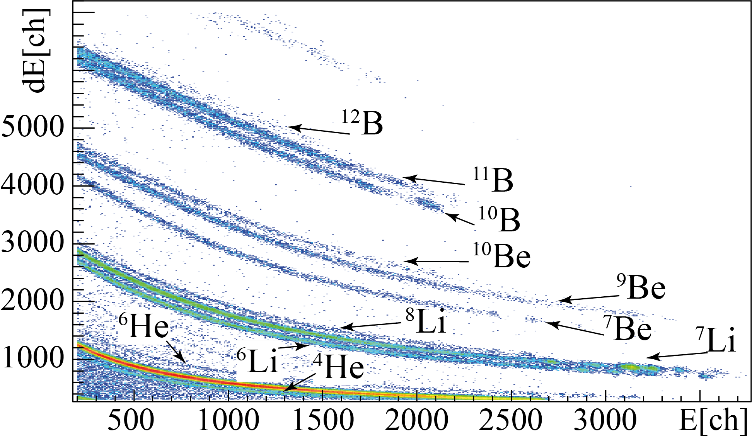}
\caption{Two-dimensional $\Delta E \times E$ spectrum of reaction products $^{7}$Li+$^{10}$B.}
\label{fig:02}
\end{figure*}

The angular distance $\Updelta \Uptheta_\text{LAB}$ between neighboring telescopes of the first and second groups is $\Updelta \Uptheta_\text{LAB} = 3.2^\circ$ and $1.7^\circ$, respectively. The $\Uptheta_\text{LAB}$ telescope angles were varied with an accuracy of $\pm 0.50^\circ$. The energy resolution of the $\Delta E$ and $E$ detectors of the telescopes was no worse than $\Delta E$ = 60 keV ($^{226}$Ra $\alpha$-source).Fig. \ref{fig:02} shows the two-dimensional $\Delta E \times E$ spectrum of the $^{7}$Li+$^{10}$B reaction  products in $\Delta E - E$ telescope (thickness $\Delta E$ -- 30 $\upmu$m and $E$ -- 800 $\upmu$m) at $\Uptheta_\text{LAB} = 22.3^\circ$.

\begin{figure*}[htb]
\includegraphics
  [width=0.9\hsize]
  {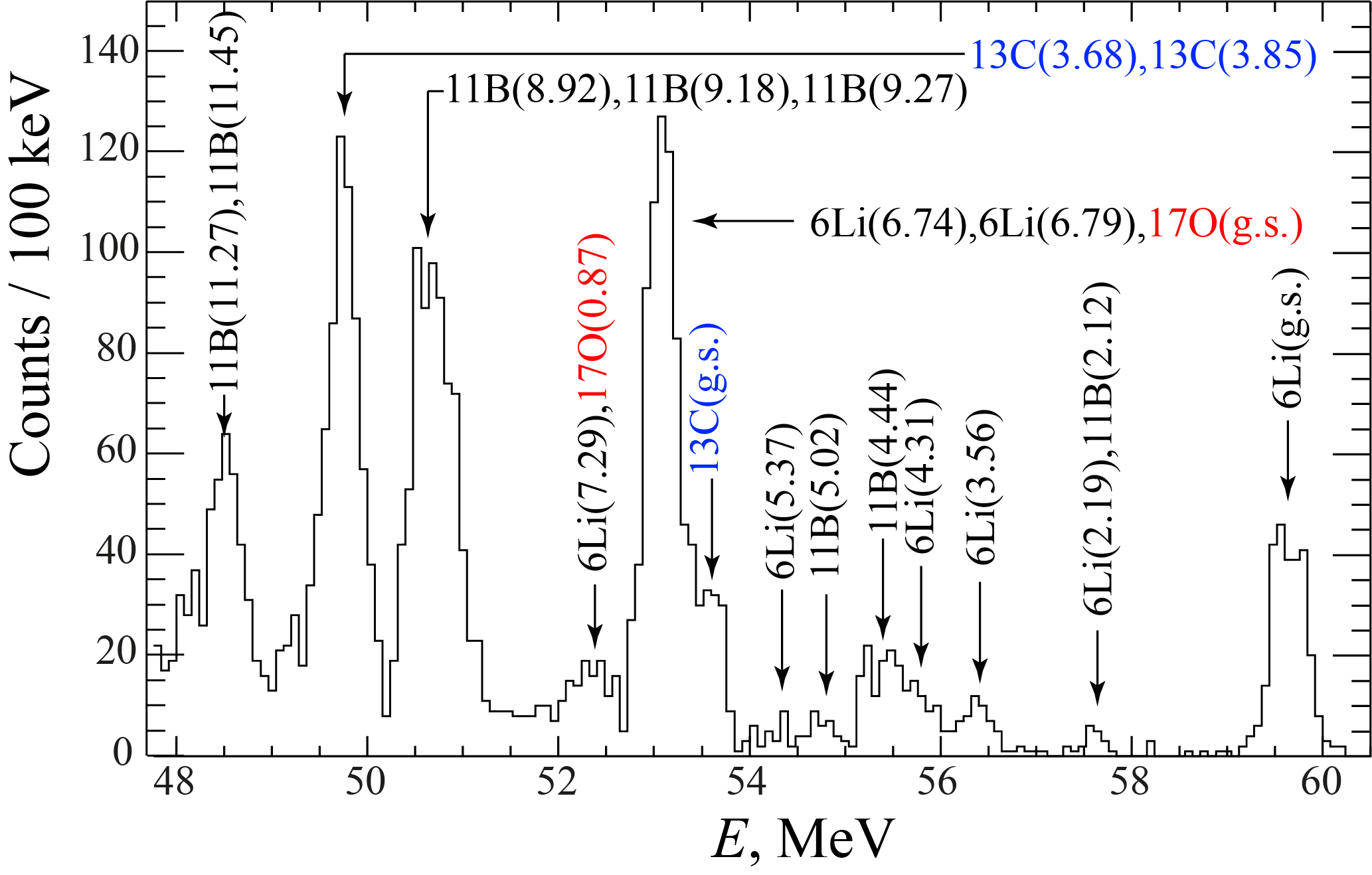}
\caption{Energy ($\Delta E + E$) spectrum of $^{6}$Li recorded at $\Uptheta_\text{LAB}$ = $10^\circ$. Black color indicates the $^{6}$Li peaks from the reaction on the $^{10}$B target, blue and red indicate the $^{6}$Li peaks obtained in reactions on $^{12}$C and $^{16}$O impurities, respectively}
\label{fig:03}
\end{figure*}

\begin{figure*}[htb]
\includegraphics
  [width=0.8\hsize]
  {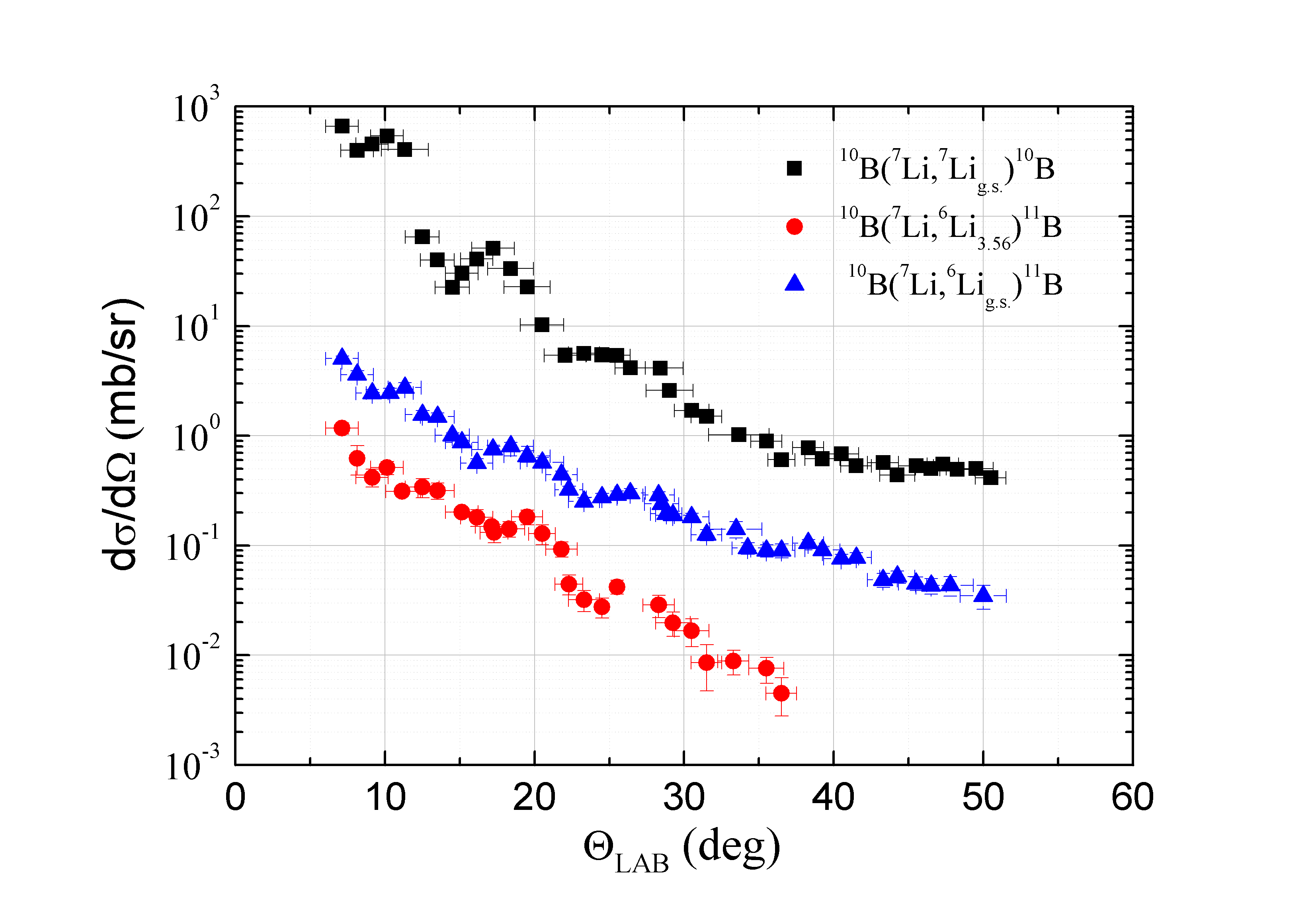}
\caption{
Angular distributions of differential cross sections of elastic scattering of ($^{7}$Li+$^{10}$B) and reaction channels of transfer $^{10}$B($^{7}$Li,$^{6}$Li)$^{11}$B to the ground $^{6}$Li and excited $^{6}$Li($J^\pi$ = 0$^+$, $T$ = 1, $E$ = 3.56 MeV) states of $^{6}$Li. Black squares correspond to elastic scattering of $^{7}$Li+$^{10}$B, blue triangles to the ground state of $^{6}$Li$_\text{g.s.}$, red circles to the excited state of $^{6}$Li($J^\pi$ = 0$^+$, $T$ = 1, $E$ = 3.56 MeV). 
}
\label{fig:04}
\end{figure*}

Fig. \ref{fig:03} shows the energy spectrum of $^{6}$Li recorded by the telescope located at $\Uptheta_\text{LAB}$ = $10^\circ$. The total ($\Delta E + E$) energy of ionization losses of $^{6}$Li in the $\Delta E$ and $E$ detectors of the telescope in MeV units is plotted along the abscissa axis. The number of events per 100 keV of the abscissa scale is plotted along the ordinate axis.

In Fig. \ref{fig:03} the positions of the peaks of the ground and excited states of the nuclei $^{6}$Li, $^{11}$B, as well as $^{13}$C, $^{17}$O, which are the products of reactions on the nuclei $^{12}$C, $^{16}$O, which constitute impurities in the target, are indicated by vertical arrows with corresponding caption.

The differential cross sections of elastic scattering of $^{7}$Li on the $^{10}$B target at the beam energy $E_\text{LAB}$=58 MeV are shown in Fig. \ref{fig:04} by filled squares.

The differential cross sections of elastic scattering of ($^{7}$Li+$^{10}$B) and the reaction channels of transfer of $^{10}$B($^{7}$Li,$^{6}$Li)$^{11}$B to the ground and excited $^{6}$Li($J^\pi$ = 0$^+$, $T$ = 1, $E$ = 3.56 MeV) states of the $^{6}$Li nucleus are shown in Fig. \ref{fig:04} by filled triangles and circles, respectively.

\section{Theory}

Direct nuclear reactions in most cases are peripheral reactions. The largest contribution from the direct reaction to the angular distribution occurs in the forward region of the angles, forming the so-called “main maximum”, which contains information about the interactions and structure of the participating nuclei. When analyzing angular distributions in the region where the main maximum of the direct transfer reaction can be observed, it seems quite adequate to use the “Distorted Wave Born Approximation” (DWBA) method \cite{Satchler1983}. The use of various structural models within the framework of this method makes it possible to obtain from such analysis information about features of the distribution of matter in the states of nuclei under study and quantities characterizing the interaction in the decay vertex $^{7}$Li $\rightarrow$ $^{6}$Li+$n$.

In DWBA, for the reaction A(a,b)B caused by the interaction V, we have the expression for the transition amplitude (see, for example, \cite{Satchler1983}):

\begin{widetext}
\begin{equation}
\label{eq:01}
    T_\text{DWBA} = \int u_b^{(-)^{*}} (k_b, r_b) \langle\Psi_b \Psi_B || V || \Psi_a \Psi_A\rangle u_a^{(+)} (k_a, r_a) dr_a dr_b
\end{equation}
\end{widetext}

$u_i$  – “distorted waves” – wave functions that describe elastic scattering in the input and output channels of the reaction, satisfying the corresponding asymptotic conditions at large distances.

Elastic scattering in this case involves the loss of flux through many other reaction channels (including the compound nucleus), none of which is the most important. These channels are represented on average using a complex optical potential (OP), which is here constructed within the phenomenological approach, using the usual Woods-Saxon parameterization:


\begin{widetext}
\begin{equation}
\begin{split}
    \label{eq:02}
    U(r) &= -V f(x_\text{V}) - i W_\text{S} f(x_\text{S}) + i 4W_\text{D} \cfrac{df(x_\text{D})}{dx_\text{D}} + V_\text{C}(r) \\
    f(x) = (e^{x} &+1)^{-1} \qquad x_i = (r_i - R_i)/a_i \qquad R_i = r_i A^{1/3} \qquad (i = \text{V, S, D})
\end{split} 
\end{equation}
\end{widetext}

The Coulomb potential $V_\text{C}(r)$ is modeled by the potential of interaction of a point charge $Z_{ae}$ with a uniformly charged sphere having a charge $Z_{Ae}$ and a radius $R_\text{C}$, which can be estimated through the root-mean-square charge radii of the nuclei. When nuclear spins are not zero, a spin-orbit potential is added:

\begin{widetext}
\begin{equation}
\label{eq:03}
    V_\text{SO}(r) =(V_\text{SO} + i W_\text{SO}) \left( \cfrac{h}{m_{\pi}c}\right)^{2} \cfrac{1}{r} \cfrac{df(x_\text{SO})}{dr} (L \cdot I_{a})
\end{equation}
\end{widetext}

The parameters are selected to describe the observed elastic scattering in a given channel at the appropriate energy, if such data are available. To be confident in the correctness of the description of elastic scattering in cases where such data are not available, additional empirical information is used on the energy and mass systematic of integral characteristics, for example, reaction cross sections and volume integrals of OP components. As initial one the parameters of the so-called “global potentials” can be used. These parameters as the functions of energy, charges and mass numbers were determined as a result of the analysis of a very large set of experimental data for a given projectile nucleus in a wide range of energies and target nuclei, or for a narrower mass region, for example, $1p$-shell nuclei.

In the DWBA amplitude \eqref{eq:01}, the matrix element is the overlap integral of the interaction $V$ and the wave functions of the interacting nuclei $(a, A)$ and product nuclei $(b, B)$ that contains all the structural information. Separation of radial and angular variables through expansion in transferred angular momentum (see, for example, \cite{Satchler1983}) identifies the product of the radial parts of the overlap integrals for the so-called “light” ($a = b + n$) and “heavy” ($B = A + n$) ) systems, often called the corresponding “cluster form factors” or “reaction form factors” $I^{ab}_{lsj}(r)$ and $I^{BA}_{lsj}(r)$, where $lsj$ are the transferred orbital, spin and total momentum and $r$ is the distance between the transferred particle $n$ and the core $A(b)$. Note that in the theory, the spectroscopic factors (SF) $S^{ab}_{lsj}$ and $S^{BA}_{lsj}$  are determined as norm of corresponding form factor. 

It is known (see, for example, \cite{Bang1974ru}) that form factors can be represented as solutions to the inhomogeneous equation:

\begin{equation}
\label{eq:04}
    \left(T_r - V^{0}_{lsj}(r) - \epsilon_{lsj}\right) I_{lsj}(r) = P(r)
\end{equation}

where $T_r$ is the operator of the kinetic energy of the relative motion of the transferred particle and the core, $V^{0}_{lsj}(r)$ is the “self-consistent field” of interaction between the particle and the core, which alone serves as the source of this field, $\epsilon_{lsj}$ is the binding energy of the transferred particle in a given state of the nucleus $B(a)$. The right side of equation \eqref{eq:04} is an integral operator, including residual interactions mixing different configurations and taking into account the Pauli principle for the transferred particle relative to the core.

There are various ways to determine form factors.  Direct one is calculation of overlap integrals using model nuclear wave functions. It is widely assumed that the form factor of the transfer reaction is identical to the wave function of the single-particle neutron state in the mean field potential of the $^{7}$Li nucleus, which corresponds to the solution of the homogeneous equation \eqref{eq:04} (with the right-hand side neglected). Other approaches involve an approximate solution of equation \eqref{eq:04}.

Here, to solve this equation, some phenomenological approach is used. The right side of equation \eqref{eq:04} is replaced by some local operator acting directly on the form factor $P(r) = \Delta V_{lsj}(r)I_{lsj}(r)$. Moving this to the left side we get a homogeneous equation with some effective potential.

\begin{gather}
\label{eq:05}
    V_{lsj}(r) = V^{0}_{lsj}(r) + \Delta V_{lsj}(r) \\
    \label{eq:06}
    (T_r - V_{lsj}(r) - \epsilon_{lsj}) I_{lsj}(r) = 0
\end{gather}

Due to the short action of these potentials, the boundary conditions at large distances are determined only by the binding energy $\epsilon_{lsj}$, i.e. the asymptotic of the form factor is expressed by the spherical Hankel function (or the Whittaker function in the case of a charged particle)

\begin{equation}
\label{eq:07}
    I_{lsj}(r \rightarrow \infty) = N^{1/2}C_{lsj} \, \kappa \, h_l(i\kappa r)
\end{equation}

$\kappa^2 = 2 \mu \epsilon_{lsj}/\hbar^{2}$, $\mu$ -- reduced mass, $N$ -- coefficient taking into account the antisymmetrization of wave functions. The value of $N C_{lsj}$ is called the asymptotic normalization coefficient (ANC), which is associated with the nuclear vertex constant (NVC) $G_{lj}$, which characterizes the interaction at the vertex of the decay $A \rightarrow n + (A-1)$:

\begin{equation}
\label{eq:08}
    G_{lj}^2 = \pi(\hbar/\mu c)^2 N C_{lj}^2 .
\end{equation}

Due to the fact that direct nuclear reactions in most cases are peripheral reactions, the angular distributions are sensitive only to the surface (asymptotic) part of the reaction form factor. This allows us to obtain empirical values of NVC and ANC from the description of the main peak of the experimental angular distribution. However, this also prevents us from obtaining the SF, the determination of which requires knowledge of the form factor for all $r$. The approach based on a combination of the concepts of the distorted wave method and the dispersion theory of direct nuclear reactions, allows one to identify purely peripheral processes that occur with the dominance of the pole mechanism of nucleon transfer. Such reactions are the main suppliers of reliable spectroscopic information (For more details, see, for example \cite{Blokhintsev1977, Goncharov1982}).

Here, to determine the form factor, ANC and NVC, approximate equation \eqref{eq:06} was solved for each configuration $lsj$ with model potential $V_{lsj}(r)$ in a simple 3-parameter Woods-Saxon form.

\begin{equation}
\label{eq:09}
    V_{lsj}(r) = V(e^{x}+1)^{-1}, \qquad x=(r-R)/a 
\end{equation}

The parameters $R$ and $a$ of the model potential $V_{lsj}(r)$ are selected to best describe the shape of the angular distribution of the reaction at least near the region of the main maximum at the forward angles, and its depth $V$ is determined for given $R$ and $a$ from the binding energy $\epsilon_{lsj}$ using the well depth fitting procedure. If the solution to equation \eqref{eq:06} is determined normalized to 1, the ANC (and, accordingly, the NVC) is determined from normalization to the absolute value of the experimental cross section. Obviously, the accuracy of determining these quantities is due to measurement error of the absolute value of the experimental cross section. In our case, we estimate this error to be $\sim$15\%.

As a result of such an analysis, we obtain radial dependences for the form factors and the ANC value, which can be compared both with the results of available theoretical calculations with model wave functions of nuclei, and with empirical values obtained from the analysis of other reactions. In particular, by comparing the form factors for different states of the nucleus of interest to us, we can obtain indirect information about the difference in the sizes of the nucleus in these states (see below).

The differential cross sections of elastic scattering and reaction were calculated using the code FRESCO \cite{Thompson1988}, with coherent consideration of all allowed combinations of transferred angular momentum and spins $lsj$.

\section{Results and analysis}

\begin{figure*}[htb]
\includegraphics
  [width=0.75\hsize]
  {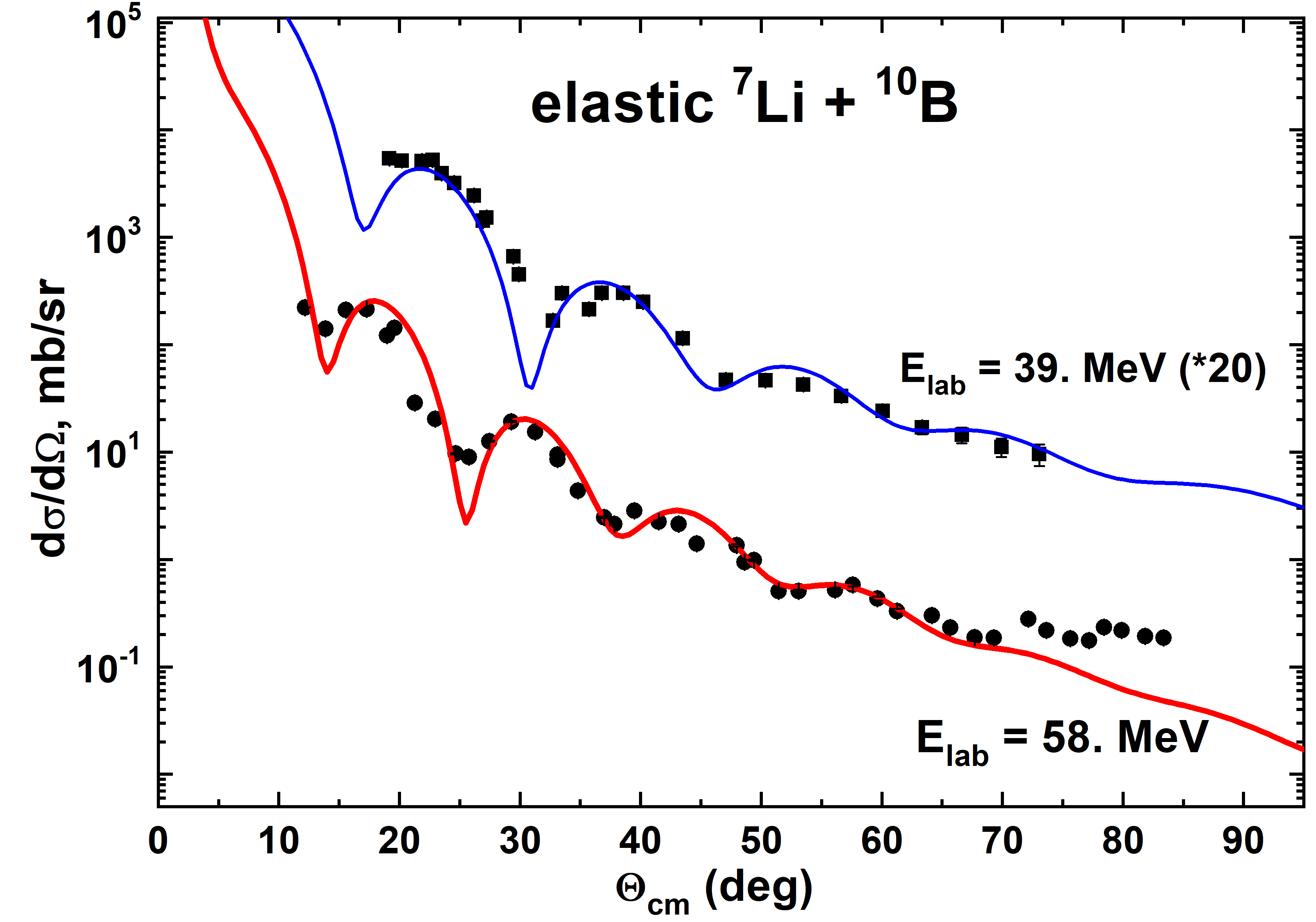}
\caption{
Differential cross sections of elastic scattering $^{7}$Li+$^{10}$B at energies of 58 and 39 MeV are presented. Black circles and squares are experimental data, solid lines are calculations with the obtained optical potentials
}
\label{fig:05}
\end{figure*}

\begin{figure*}[htb]
\includegraphics
  [width=0.75\hsize]
  {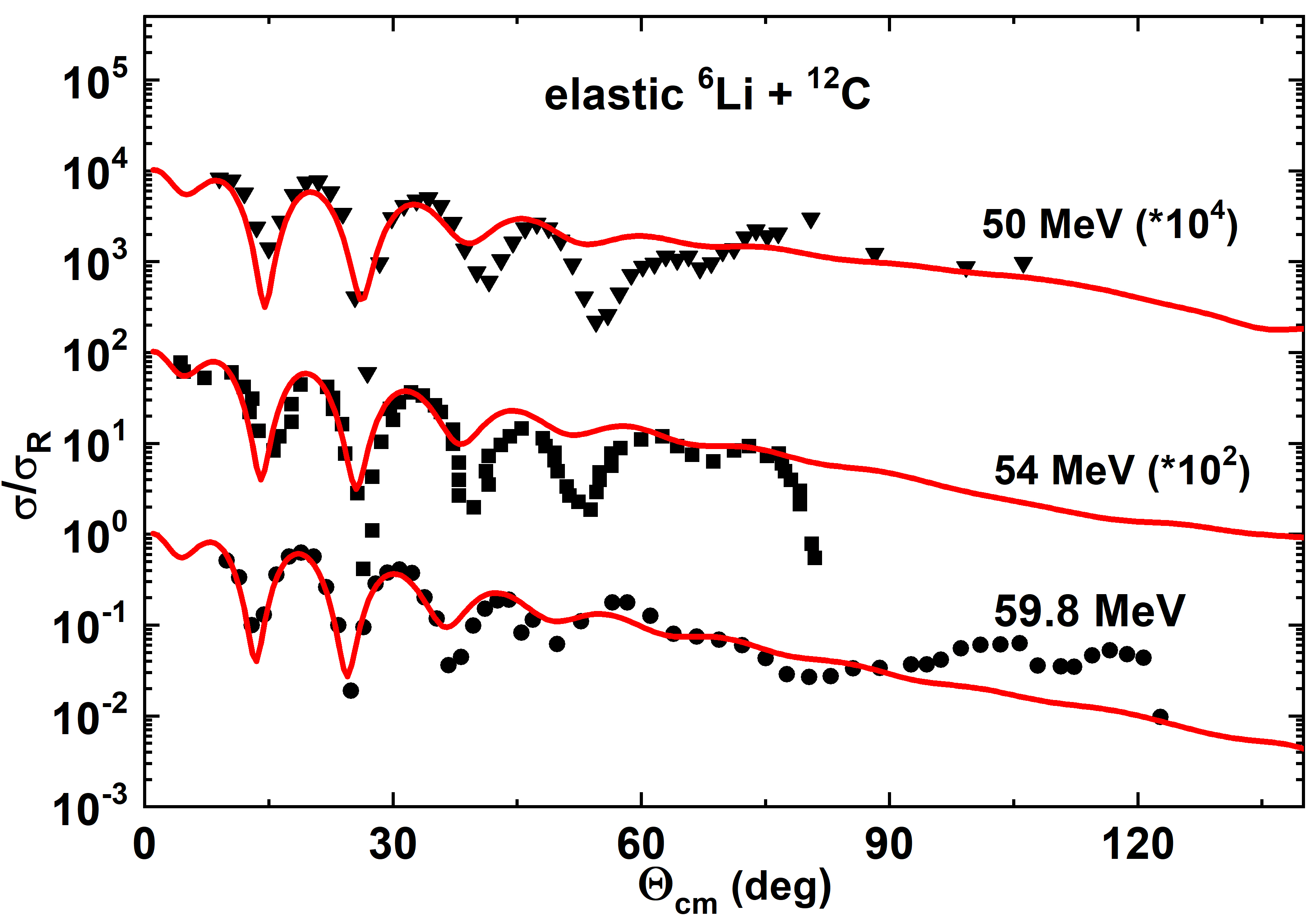}
\caption{
Differential cross sections of elastic scattering $^{6}$Li+$^{12}$C at energies of 50, 54 and 59.8 MeV are present. Black circles, triangles and squares are experimental data \cite{Guzhovskij1977, Sherman1976, Kerr1995}, solid lines are calculations with the obtained optical potentials
}
\label{fig:06}
\end{figure*}

\newpage

\begin{figure*}[htb]
\includegraphics
  [width=0.75\hsize]
  {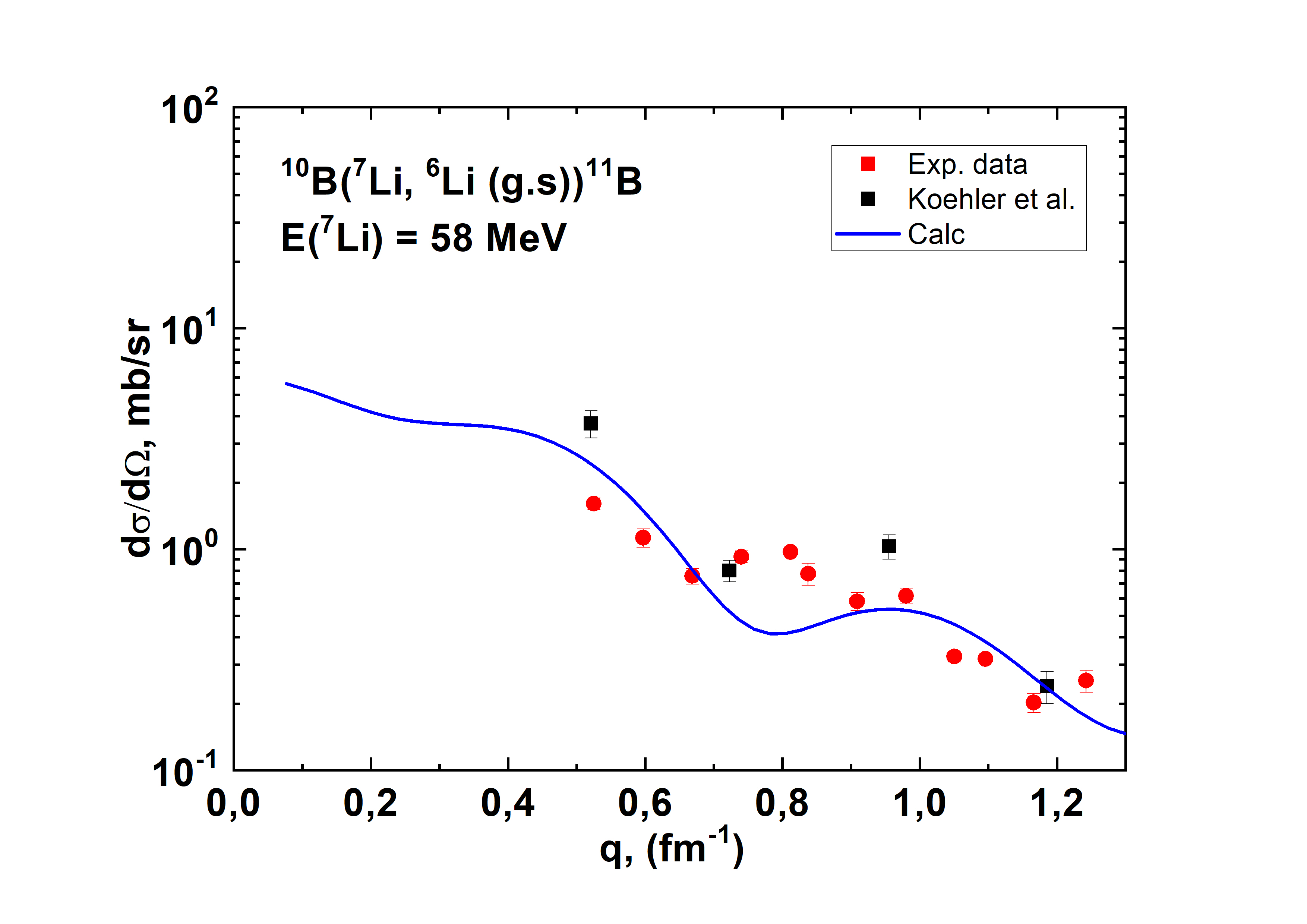}
\caption{
Differential cross sections of the reaction $^{10}$B($^{7}$Li,$^{6}$Li)$^{11}$B for ground state of $^{6}$Li. Red circles – our data at 58 MeV, squares – data at 24 MeV from \cite{Kohler1977}, solid line – the DWBA calculations.
}
\label{fig:07}
\end{figure*}

The optical potential in the input channel of $^{7}$Li+$^{10}$B was constructed based on the analysis of our elastic scattering data at 58 MeV and the data from \cite{Etchegoyen1988} at 39 MeV. In both cases, the same sets of geometric parameters (effective radii and diffuseness) and parameters of the spin-orbit and Coulomb potentials were used:

\begin{widetext}
    \begin{equation}
    \label{eq:10}
        \begin{gathered}
             r_R = 0.55 \, \text{fm}, \quad  a_R = 0.86 \, \text{fm}, \quad   r_S = 0.88 \, \text{fm}, \quad  a_S = 0.30 \,\text{fm}, \quad   r_D = 0.65 \, \text{fm}, \quad a_D = 0.95 \, \text{fm}, 
             \\
             r_C = 0.60 \, \text{fm}, \quad V_{SO} = 6.0 \, \text{MeV}, \quad   r_{SO} = 0.55 \, \text{fm}, \quad   a_{SO} = 0.86 \, \text{fm}
        \end{gathered}
    \end{equation}
\end{widetext}

Here: 

\begin{equation}
\label{eq:11}
    R_i = r_i (A_p^{1/3} + A_{t}^{1/3}), \quad (i = R, S, D, C, SO).
\end{equation}

In this case, reasonable values and the trend of the energy dependence were obtained for the force parameters (potential depths), volume integrals and reaction cross sections (see Table \ref{tab:01}). The description of the angular distributions is presented in Figure \ref{fig:05}.

\begin{table}[h]
    \centering
    \caption{Optical potentials in Woods-Saxon form \eqref{eq:02} for $^{7}$Li+$^{10}$B}
    \label{tab:01}
    \begin{tabular*}{8.5cm} {@{\extracolsep{\fill} } *{7}{c}}
    \toprule
        $E_\text{LAB},$ & $V_r,$ & $W_S,$ & $W_D,$ & $-J_V,$ & $-J_W,$ & $\sigma_r,$  \\
        MeV & MeV & MeV & MeV & MeV fm$^3$ & MeV fm$^3$ & mb
        \\
        \midrule
        58.0 & 250. & 10.0 & 22.0 & 415. & 177. & 1468.
        \\
        39.0 & 290. & 7.0 & 16.0 & 481. & 128. & 1434
        \\
        \bottomrule
    \end{tabular*}
\end{table}

Unfortunately, there are no experimental data on elastic scattering for the output channels $^{6}$Li+$^{11}$B in the energy range of interest to us. However, there are data on scattering $^{6}$Li+$^{12}$C \cite{Guzhovskij1977, Sherman1976, Kerr1995} at 50, 54 and 59.8 MeV that can be relied upon. We tried to construct a kind of regional potential for the scattering of lithium nuclei on 1p-shell nuclei by fixing the geometric parameters \eqref{eq:10} and using these data. Table \ref{tab:02} presents the energy dependent parameters and integral characteristics obtained from the analysis of $^{6}$Li+$^{12}$C data. Figure \ref{fig:06} shows the quality of description of the corresponding experimental angular distributions.

\begin{table}[h]
    \centering
    \caption{Optical potentials in Woods-Saxon form \eqref{eq:02} for $^{6}$Li+$^{12}$C}
    \label{tab:02}
    \begin{tabular*}{8.5cm} {@{\extracolsep{\fill} } *{7}{c}}
    \toprule
        $E_\text{LAB},$ & $V_r,$ & $W_S,$ & $W_D,$ & $-J_V,$ & $-J_W,$ & $\sigma_r,$  \\
        MeV & MeV & MeV & MeV & MeV fm$^3$ & MeV fm$^3$ & mb
        \\
        \midrule
        59.8 & 255. & 14.0 & 19.5 & 419. & 170. & 1392.
        \\
        54.0 & 270. & 12.0 & 18.0 & 444. & 154. & 1385.
        \\
        50.0 & 300. & 10.0 & 16.0 & 493. & 135. & 1372.
        \\
        \bottomrule
    \end{tabular*}
\end{table}

Note that using a minimum number of variable parameters, it was possible to obtain the correct energy behavior of the potential characteristics and a fairly good description of the data in the region of forward angles. 

Using the geometric parameters fixed in this way, taking into account the dependence on the mass number and the energy trend of the force parameters, we evaluated the parameters for the $^{6}$Li+$^{11}$B output channels at real energies corresponding to the yield of the $^{6}$Li nucleus in the ground and excited states. The parameters used for reaction calculations are given in Table \ref{tab:03}.

\begin{widetext}
    \begin{table}[h]
    \renewcommand{\arraystretch}{1.5}
    \centering
    \caption{Optical potentials in Woods-Saxon form \eqref{eq:02} for $^{6}$Li+$^{12}$C}
    \label{tab:03}
    \begin{tabular*}{17.5cm}{@{\extracolsep{\fill} } *{8}{c}}
    \toprule
         & $E_\text{LAB},$ & $V_r,$ & $W_S,$ & $W_D,$ & $-J_V,$ & $-J_W,$ & $\sigma_r,$  \\
        & MeV & MeV & MeV & MeV & MeV fm$^3$ & MeV fm$^3$ & mb
        \\
        \midrule
        $^{6}$Li$_\text{(g.s.)}$ + $^{11}$B & 59.2 & 285. & 13.0 & 17.0 & 496. & 159. & 1365.
        \\
        $^{6}$Li*(3.56) + $^{11}$B & 53.7 & 285. & 12.0 & 16.0 & 496. & 149. & 1362.
        \\
        \bottomrule
    \end{tabular*}
\end{table}
\end{widetext}

Using the obtained optical potentials, we analyzed the measured differential cross sections of the reaction in two stages. First, to fix the form factor \{$^{11}$B(3/2$^{-}$, $T$=1/2, g.s.), n$^{10}$B(3$^{+}$, $T$=0, g.s.)\}, we used the empirical form factor {$^{7}$Li(3/2$^{-}$, $T$ = 1/2, g.s.), n$^{6}$Li(1$^{+}$, $T$ = 0, g.s.)} recently obtained by a similar approach from the analysis of the $^{7}$Li($d,t$)$^{6}$Li reaction at a deuteron energy of 14.5 MeV \cite{Demyanova2024}. This form factor and its ANC are in good agreement with both the empirical form factor previously obtained from the analysis of the $^{7}$Li($d,t$)$^{6}$Li reaction at a deuteron energy of 18 MeV \cite{Gulamov1995} and the theoretical calculations \cite{Timofeyuk2010} based on an approximate solution of the equation \eqref{eq:04} for the form factor by the Green's function method using a truncated shell basis and various models of effective nucleon-nucleon potentials. Thus we have defined and used the values of the obtained model potential parameters and ANC for the form factor  \{$^{11}$B(3/2$^{-}$, $T$=1/2, g.s.), n$^{10}$B(3$^{+}$, $T$=0, g.s.)\}.

We also included in the consideration the data in the region of forward angles obtained in \cite{Kohler1977} at energy of 24 MeV for the ground state of $^{6}$Li. For ease of comparison, Figure \ref{fig:07} shows our data and the data from \cite{Kohler1977} and the DWBA calculation as a function of the transferred momentum in the region of forward angles.

\begin{figure*}[htb]
\includegraphics
  [width=0.75\hsize]
  {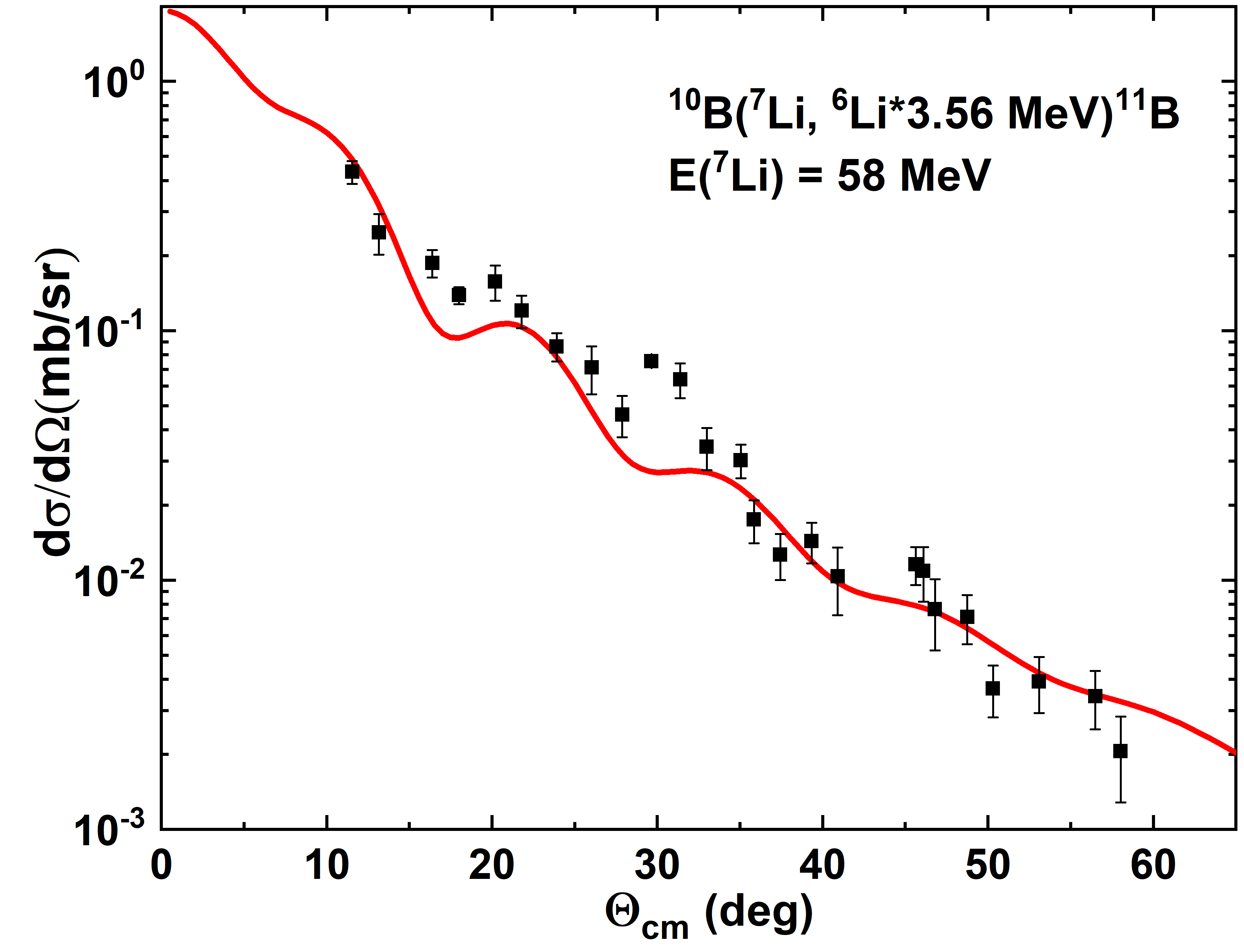}
\caption{
Differential cross sections of the reaction $^{10}$B($^{7}$Li,$^{6}$Li*)$^{11}$B at 58 MeV with excitation of the 3.56 MeV state of $^{6}$Li. Black circles are experimental data, solid lines are the DWBA calculations
}
\label{fig:08}
\end{figure*}

Having thus fixed the form factor \{$^{11}$B(3/2$^{-}$, $T$=1/2, g.s.), n$^{10}$B(3$^{+}$, $T$=0, g.s.)\}, we determined the parameters of the model potential, the radial form and the ANC of the form factor \{$^{7}$Li(3/2$^{-}$, $T$=1/2, g.s.), n$^{6}$Li(0$^{+}$, $T$=1, 3.56)\} from the analysis of the measured differential cross sections in the channel with yield of the $^{6}$Li in the excited state $0^{+}_{1}$(3.56 MeV). Figure \ref{fig:08} shows the angular distributions of the reaction (calculated and experimental) for the excited 3.56 MeV state of the $^{6}$Li nucleus. 

The values of the obtained model potential parameters and ANC and their comparison with the results of analysis and theoretical calculations of other authors are presented in the Table \ref{tab:04}. 

\begin{widetext}
    \begin{table}[!h]
    \renewcommand{\arraystretch}{1.5}
    \centering
    \caption{The values of the obtained model potential parameters and ANC}
    \label{tab:04}
    \begin{tabular*}{17.5cm} {@{\extracolsep{\fill} } *{9}{c}}
    \toprule
         A+1($I^\pi, T, E^{*}$), nA($I^\pi, T, E^{*}$) & $(l,j)$ & $-V$ & $R$ & $a$ & $<r^2>^{1/2}$ & $NC^2$ & $NC^2_\text{theor}$ & $NC^2_\text{exp}$
        \\
        & & MeV & fm & fm & fm & fm$^{-1}$ & fm$^{-1}$ & fm$^{-1}$
        \\
        \midrule
        $^{7}$Li(3/2$^-$, 1/2, g.s.), n$^{6}$Li(1$^+$, 0, g.s.) & (1, 3/2) + (1, 1/2) & 60.2 & 2.36 & 0.9 & 3.23 & 3.05(46)\cite{Demyanova2024} & 2.94\cite{Timofeyuk2010} & 3.17(53)
        \cite{Gulamov1995}  
        \\
        $^{7}$Li(3/2$^-$, 1/2,  g.s.), n$^{6}$Li(0$^+$, 1, 3.56) & (1, 3/2) & 66.7 & 2.18 & 1.6 & 3.50 & 2.76(39) & 2.47\cite{Timofeyuk2010} & 2.91(35) \cite{Gulamov1995}
        \\
        & & & & & & & & 3.16(47) \cite{Demyanova2024}
        \\
        $^{11}$B(3/2$^-$, 1/2, g.s.), n$^{10}$B(3$^+$, 0, g.s.) & (1, 3/2) & 43.3 & 3.34 & 0.5 & 2.89 & 12.4(1.9) & 13.1\cite{Timofeyuk2010} & 31.4(3.0) \cite{Gulamov1995}
        \\
        \bottomrule
    \end{tabular*}
\end{table}
\end{widetext}

\begin{figure*}[htb]
\includegraphics
  [width=0.8\hsize]
  {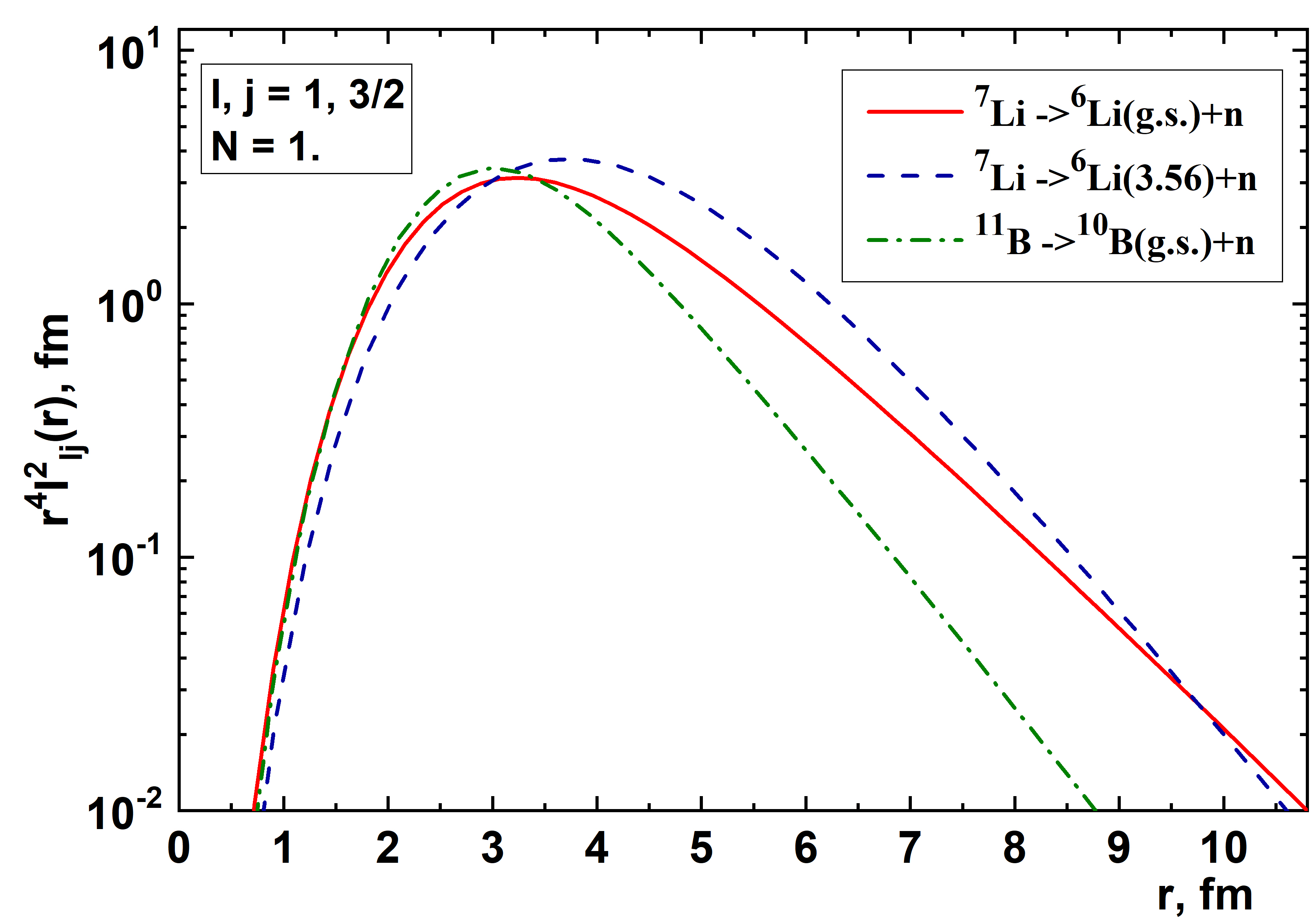}
\caption{
Comparison of the radial dependences of the values $r^{4}I_{lj}^2(r)$ for all three form factors
}
\label{fig:09}
\end{figure*}

\newpage

We note that the root-mean-square radius of the form factor \{$^{7}$Li(3/2$^{-}$, $T$=1/2, g.s.), n$^{6}$Li(0$^{+}$, $T$=1, 3.56)\} turned out to be larger than that for the ground state of $^{6}$Li. Note that this result is in some contradiction to results of ab initio calculations where radii of the $^{6}$Li and $^{6}$He g.s. are practically equal to the radius of the 3.56 MeV state \cite{Rodkin2023}. In \cite{Timofeyuk2010} the $rms$ radii of form factors are presented and for the excited state of $^{6}$Li this radius turned out to be even 5\% smaller, while in our case this radius is 8\% larger. In our work, empirical radial dependences for form factors were obtained, which can be compared with theoretical model form factors, but, unfortunately, such were not presented in \cite{Timofeyuk2010}. Therefore, it is difficult to explain due to what these small discrepancies arise, which can be within the uncertainties of both theoretical model calculations and experimental errors. Figure 9 shows a comparison of the radial dependences of the values $r^{4}I_{lj}^{2}(r)$, which are the integrands in calculating the mean-square radii for all three form factors normalized to 1. It is clearly seen that the main contribution to the value of these radii comes from the region from 1 to 10 fm, and in the region from 2 to 9 fm we see a significant difference in the radial dependence for the transition to the ground and excited states of $^{6}$Li. This difference corresponds to an increased value of the $rms$ radius of the form factor of the transition to the excited state. Considering that in both cases the contribution to the form factor comes from the same wave function of the ground state of the $^{7}$Li nucleus, it is possible to assert that it is the wave function of the $^{6}$Li nucleus that is responsible for this difference. However, this gives us only a qualitative estimate that the dimensions of the $^{6}$Li nucleus in the excited state are "larger" than in the ground state. The reason for this may be the presence of a halo structure. 

The radial dependence of the form factor for \{$^{11}$B, $n {}^{10}$B\} is shown in Fig. 9 as an independent result. No papers were found in the literature that contained results for the radial dependence of this form factor.

Our results within the errors are in good agreement with the ANC values calculated in \cite{Timofeyuk2010} and obtained from the analysis of the $^{7}$Li($d,t$)$^{6}$Li reaction at a deuteron energy of 18 MeV \cite{Gulamov1995}. As for the disagreement with the previous experiment $^{11}$B($d,t$)$^{10}$B \cite{Gulamov1995} of the ANC\{$^{11}$B,$^{10}$B+$n$\}, we find it difficult to comment, since \cite{Gulamov1995} does not present the details of the analysis from which the NVC was determined, that for comparison we recalculated in ANC. In particular, the optical potentials used, their origin and adequacy are not discussed at all, which naturally can affect the uncertainty of the final result.

\section{Conclusion}

$^{7}$Li elastic scattering and the lithium-induced reaction of one-nucleon transfers from $^{10}$B($^{7}$Li,$^{6}$Li)$^{11}$B {have} been measured at $E_\text{LAB}$($^{7}$Li) = 58 MeV. Experiment was done at the U-400 accelerator beam of the FLNR JINR, Dubna. Angular distributions for the ground and the 3.56 MeV excited states of the $^{6}$Li were measured in the lab angular range $\Uptheta_\text{LAB}$=($7^\circ - 50^{\circ}$) Angular distribution for the $^{6}$Li($J^\pi$ = 0$^+$, $T$ = 1, $E$ = 3.56 MeV) state are presented for the first time.

The DWBA analysis of the differential cross section of the $^{10}$B($^{7}$Li,$^{6}$Li)$^{11}$B reaction with excitation of the $^{6}$Li$_\text{g.s.}$ and the $^{6}$Li($J^\pi$ = 0$^+$, $T$ = 1, $E$ = 3.56 MeV) states was performed.  The use of DWBA in the region of the main peak of the angular distribution is quite adequate, since the reaction is direct and peripheral and allows obtaining information about the interactions and structure of the participating nuclei. As a first step of analysis, optical model potentials were obtained by fitting measured elastic scattering data and evaluating optical potential parameters for the output reaction channels. Using a minimum number of variable parameters, we have obtained the correct energy behavior of the potential characteristics and a fairly good description of the data in the region of forward angles. 

Phenomenological approach based on solving an approximate equation for the form factor was used to determine radial dependence of the reaction form factor. As reaction is peripheral, the angular distributions are sensitive only to the surface (asymptotic) part of the reaction form factor. This allowed us to obtain empirical values of ANC and empirical radial dependences of the form factors for all three decay vertices: \{$^{7}$Li(g.s.) $\rightarrow$ $n$+$^{6}$Li(g.s.)\}, \{$^{7}$Li(g.s.) $\rightarrow$ $n$+$^{6}$Li($0^+$,$T$ = 1, 3.56)\} and \{$^{11}$B(g.s.) $\rightarrow$ $n$+$^{10}$B(g.s.)\}. No papers were found in the literature that contained results for the radial dependence of form factor for \{$^{11}$B(g.s.) $\rightarrow$ $n$+$^{10}$B(g.s.)\}.

Obtained values of ANC’s for $^{6}$Li$_\text{g.s.}$ and $^{6}$Li($J^\pi$ = 0$^+$, $T$ = 1, $E$ = 3.56 MeV) states are similar to literature one. This fact confirms correctness of our DWBA analysis. 

Comparison of the radial dependences of form factors shows significant difference in the radial dependence for the transition to the ground and excited states of $^{6}$Li, which corresponds to an increased value of the $rms$ radius of the form factor of the transition to the excited state. Considering that for both cases the contribution to the form factor comes from the same wave function of the ground state of the $^{7}$Li nucleus, it is possible to assert that it is the wave function of the $^{6}$Li nucleus that is responsible for this difference. Thus, the wave function of the $^{6}$Li nucleus in the $^{6}$Li($J^\pi$ = 0$^+$, $T$ = 1, $E$ = 3.56 MeV) state has increased spatial dimension compared to the $^{6}$Li$_\text{g.s.}$. However, this gives us only a qualitative estimate that the radius of the $^{6}$Li nucleus in the 3.56 MeV state is larger than in the ground state. This result is one of the arguments in favor of a halo existence in $^{6}$Li*(3.56 MeV) state, while the question of a halo in $^{6}$Li$_\text{g.s.}$ still leaves open.


\begin{thebibliography}{99}

\bibitem{TANIHATA1985}
I.~Tanihata, H.~Hamagaki, O.~Hashimoto, {\textit{\textbf{et al.,}}}
\newblock {Physical Review Letters}, 55(24):2676--2679 (1985).
\newblock DOI: \url{https://doi.org/10.1103/PhysRevLett.55.2676}

\bibitem{Riisager2013}
K~Riisager,
\newblock {Physica Scripta}, T152:014001 (2013).
\newblock DOI: \url{https://doi.org/10.1088/0031-8949/2013/t152/014001}

\bibitem{Izosimov2016}
Igor Izosimov,
\newblock {EPJ Web of Conferences}, 107:09003 (2016).
\newblock DOI: \url{https://doi.org/10.1051/epjconf/201610709003}

\bibitem{Izosimov2017}
I.~N. Izosimov,
\newblock {Physics of Atomic Nuclei}, 80(5):867–876 (2017).
\newblock DOI: \url{https://doi.org/10.1134/S1063778817050118}

\bibitem{Izosimov2018}
I.~N. Izosimov,
\newblock {Physics of Particles and Nuclei Letters}, 15(6):621–626 (2018).
\newblock DOI: \url{https://doi.org/10.1134/S1547477118060092}

\bibitem{Izosimov2020}
Igor Izosimov,
\newblock {EPJ Web of Conferences}, 239:02003 (2020).
\newblock DOI: \url{https://doi.org/10.1051/epjconf/202023902003}

\bibitem{Sobolev2005}
Yu~Sobolev, A.~Budzanowski, E.~Bialkowski, {\textit{\textbf{et al.,}}}
\newblock {Bulletin of The Russian Academy of Sciences: Physics}, 69:1790--1795 (2005).


\bibitem{Lukyanov2008}
Konstantin Lukyanov, Elena Zemlyanaya, V.~Lukyanov, {\textit{\textbf{et al.,}}}
\newblock {Bulletin of The Russian Academy of Sciences: Physics}, 72:356--360 (2008).
\newblock DOI: \url{https://doi.org/10.3103/s11954-008-3019-7}

\bibitem{Kalpakchieva2007}
R.~Kalpakchieva, V.~A. Maslov, R.~A. Astabatian,{\textit{\textbf{et al.,}}}
\newblock {Physics of Atomic Nuclei}, 70(4):619–625 (2007).
\newblock DOI: \url{https://doi.org/10.1134/S1063778807040023}

\bibitem{CortinaGil1996}
M.D. Cortina-Gil, P.~Roussel-Chomaz, N.~Alamanos, {\textit{\textbf{et al.,}}}
\newblock {Physics Letters B}, 371(1–2):14–18 (1996).
\newblock DOI: \url{https://doi.org/10.1016/0370-2693(95)01582-5}

\bibitem{CortinaGil1998}
M.D. Cortina-Gil, A.~Pakou, N.~Alamanos, {\textit{\textbf{et al.,}}}
\newblock {Nuclear Physics A}, 641(3):263–270 (1998).
\newblock DOI: \url{https://doi.org/10.1016/s0375-9474(98)00470-9}

\bibitem{Brown1996}
J.~A. Brown, D.~Bazin, W.~Benenson, {\textit{\textbf{et al.,}}}
\newblock {Physical Review C}, 54(5):R2105–R2108 (1996).
\newblock DOI: \url{https://doi.org/10.1103/physrevc.54.r2105}

\bibitem{Li2002}
Zhihong Li, Weiping Liu, Xixiang Bai, {\textit{\textbf{et al.,}}}
\newblock {Physics Letters B}, 527(1–2):50–54 (2002).
\newblock DOI: \url{https://doi.org/10.1016/s0370-2693(02)01172-3}

\bibitem{Galanina2014}
L.~I. Galanina and N.~S. Zelenskaya.
\newblock {Physics of Atomic Nuclei}, 77(6):704–715 (2014).
\newblock DOI: \url{https://doi.org/10.1134/s106377881405007x}

\bibitem{Demyanova2018}
A~S Demyanova, A~A Ogloblin, A~N Danilov, {\textit{\textbf{et al.,}}}
\newblock {KnE Energy}, 3(1):1, (2018).
\newblock DOI: \url{https://doi.org/10.18502/ken.v3i1.1715}


\bibitem{Danilov2009}
A.~N. Danilov, T.~L. Belyaeva, A.~S. Demyanova, {\textit{\textbf{et al.,}}}
\newblock {Physical Review C}, 80(5):054603 (2009).
\newblock DOI: \url{https://doi.org/10.1103/physrevc.80.054603}

\bibitem{Givens1972}
R.W. Givens, M.K. Brussel, and A.I. Yavin.
\newblock {Nuclear Physics A}, 187(3):490–500 (1972).
\newblock DOI: \url{https://doi.org/10.1016/0375-9474(72)90674-4}

\bibitem{Rodkin2023}
D.~M. Rodkin and Yu.~M. Tchuvil’sky.
\newblock {JETP Letters}, 118(3):153–159 (2023).
\newblock DOI: \url{https://doi.org/10.1134/s0021364023602130}

\bibitem{Tanihata1985a}
I.~Tanihata, H.~Hamagaki, O.~Hashimoto, 054603
\newblock {Physical Review Letters}, 55(24):2676–2679 (1985).
\newblock DOI: \url{https://doi.org/10.1103/physrevlett.55.2676}

\bibitem{Dobrovolsky2006}
A.V. Dobrovolsky, G.D. Alkhazov, M.N. Andronenko, {\textit{\textbf{et al.,}}}
\newblock {Nuclear Physics A}, 766:1–24 (2006).
\newblock DOI: \url{https://doi.org/10.1016/j.nuclphysa.2005.11.016}

\bibitem{Etchegoyen1988}
A.~Etchegoyen, M.~C. Etchegoyen, E.~D. Izquierdo, {\textit{\textbf{et al.,}}}
\newblock {Physical Review C}, 38(5):2124–2133 (1988).
\newblock DOI: \url{https://doi.org/10.1103/physrevc.38.2124}

\bibitem{Schumacher1973}
P.~Schumacher, N.~Ueta, H.H. Duhm,{\textit{\textbf{et al.,}}}
\newblock {Nuclear Physics A}, 212(3):573–599 (1973).
\newblock DOI: \url{https://doi.org/10.1016/0375-9474(73)90824-5}

\bibitem{mesytec}
Detector readout systems.
\newblock \url{http://https://www.mesytec.com/}.
\newblock Accessed: 2024-08-15.

\bibitem{Satchler1983}
G~R Satchler.
\newblock {\em {D}irect {N}uclear {R}eactions}.
\newblock International Series of Monographs on Physics. Clarendon Press, Oxford, England, 1983.


\bibitem{Bang1974ru}
E.~Bang, V.~E. Bunakov, F.~A. Gareev {\textit{\textbf{et al.,}}}
\newblock {Fiz. Elem. Chast. Atom. Yadra}, 5:263--307, (1974).

\bibitem{Blokhintsev1977}
L.~Blokhintsev, I.~Borbely, and E.I. Dolinskii.
\newblock {Sov. J. Particles Nucl}, 8(11):1189 (1977).

\bibitem{Goncharov1982}
S.A.~Goncharov, J.~Dobesh, E.I.~Dolinskii {\textit{\textbf{et al.,}}}
\newblock {Sov. J. Nucl. Phys.}, 35(3) (1982).

\bibitem{Thompson1988}
Ian~J. Thompson.
\newblock {Computer Physics Reports}, 7(4):167–212 (1988).
\newblock DOI: \url{https://doi.org/10.1016/0167-7977(88)90005-6}

\bibitem{Guzhovskij1977}
B.Ja. Guzhovskij, S.N. Abramovich, B.M. Dzuba {\textit{\textbf{et al.,}}}
\newblock {Problemy Yadernoj Fiziki i Kosmicheskikh Luchejs}, 7:41 (1977).

\bibitem{Sherman1976}
J.~D. Sherman, E.~R. Flynn, Nelson Stein {\textit{\textbf{et al.,}}}
\newblock {Physical Review C}, 13(6):2122–2126 (1976).
\newblock DOI: \url{https://doi.org/10.1103/PhysRevC.13.2122}

\bibitem{Kerr1995}
P.~L. Kerr, K.~W. Kemper, P.~V. Green {\textit{\textbf{et al.,}}}
\newblock {Physical Review C}, 52(4):1924–1933 (1995).
\newblock DOI: \url{https://doi.org/10.1103/physrevc.52.1924}

\bibitem{Demyanova2024}
A.S.~Demyanova et~al.
\newblock {Dimensions of $^{6}${L}i in low-lying states.}
\newblock In {\em Proceedings of the LXXIV International conference "NUCLEUS-2024, Fundamental problems and applications"} Dubna, July 1-5, page 155 (2024).

\bibitem{Gulamov1995}
I.~Gulamov, Akram Mukhamedzhanov, and G.~Nie,
\newblock {Physics of Atomic Nuclei}, 58: 1689--1695 (1995).

\bibitem{Timofeyuk2010}
N.~K. Timofeyuk.
\newblock {Physical Review C}, 81(6):064306 (2010).
\newblock DOI: \url{https://doi.org/10.1103/physrevc.81.064306}

\bibitem{Kohler1977}
W.~Kohler, G.~Gruber, A.~Steinhauser {\textit{\textbf{et al.,}}}
\newblock {Nuclear Physics A}, 290(1):233–252 (1977).
\newblock DOI: \url{https://doi.org/10.1016/0375-9474(77)90677-7}

\end{thebibliography}
\end{document}